\documentclass[useAMS,usenatbib]{mnras}

\usepackage{graphicx}	
\usepackage{amsmath}	
\usepackage{amssymb}	
\usepackage{natbib}

\title[Cosmic rays from a black hole in NGC 253]{Cosmic rays from the nearby starburst galaxy NGC 253: the effect of a low luminosity active galactic nucleus}

\author[Guti\'errez et al.]{
E. M. Guti\'errez,$^{1}$\thanks{E-mail: emgutierrez@iar.unlp.edu.ar}
G. E. Romero$^{1,2}$
F. L. Vieyro$^{1,2}$
\\
$^{1}$Instituto Argentino de Radioastronom\'ia (IAR, CCT La Plata, CONICET/CIC), C.C.5, (1894) Villa Elisa, Buenos Aires, Argentina\\
$^{2}$Facultad de Ciencias Astron\'omicas y Geof\'isicas, Universidad Nacional de La Plata, Paseo del Bosque s/n, 1900 La Plata, Buenos Aires, Argentina
}

\date{Accepted XXX. Received YYY; in original form ZZZ}

\pubyear{2019}

\begin{document}
\label{firstpage}
\pagerange{\pageref{firstpage}--\pageref{lastpage}}
\maketitle

\begin{abstract}
NGC 253 is a nearby starburst galaxy in the Sculptor group located at a distance of $\sim 3.5$ Mpc that has been suggested by some authors as a potential site for cosmic-ray acceleration up to ultra-high energies. Its nuclear region is heavily obscured by gas and dust, which prevents establishing whether or not the galaxy harbours a supermassive black hole coexisting with the starburst. Some sources have been proposed in the literature as candidates for an active nucleus.
In this work, we aim at determining the implications that the presence of a supermassive black hole at the nucleus of NGC 253 might have on cosmic ray acceleration.
With this aim, we model the accretion flow on to the putative active nucleus, and we evaluate the feasibility of particle acceleration by the black hole dynamo mechanism. As a by-product, we explore the potential contribution from non-thermal particles in the accretion flow to the high-energy emission of the galaxy.
We found that in the three most plausible nucleus candidates, the emission of the accretion flow would inhibit the black hole dynamo mechanism. To rule out completely the influence that a putative nucleus in NGC 253 might have in cosmic-ray acceleration, a better clarification concerning the true nature of the nucleus is needed.
\end{abstract}




\section{Introduction}

Starburst galaxies are characterised by an ongoing enhanced rate of star formation that results in the existence of a large number of hot, massive stars and stellar remnants concentrated into a small region. The many supernova remnants in this region give rise to a high density of locally accelerated cosmic rays that produce non-thermal electromagnetic radiation from radio up to gamma rays  (e.g. \citealt{paglione1996, Bykov2001,romero2003b,domingo2005,rephaeli2010}). The two nearest starburst galaxies, NGC 253 and M82, have been both detected in the high-energy and the very high-energy gamma-ray bands \citep{acero2009,abdo2010, Ackermann2012} confirming the early theoretical predictions. The gamma-ray emission is usually interpreted as the effect of cosmic ray interactions with the dense gas of the disk, although a large-scale contribution from the galactic superwind might also be present \citep{romero2018}. 

The existence of a superwind caused by the collective effects of stars and supernovae in starburst galaxies was suggested long ago by \cite{chevalier1985}. Today, the presence of a superwind can be directly established by line and continuum observations in nearby galaxies such as NGC 253 (see e.g. \citealt{veilleux2005} and references therein). The reverse shock of these superwinds has been proposed as a possible site for the acceleration of cosmic rays up to ultra-high energies \citep{anchordoqui1999,anchordoqui2018}. The idea is attractive because the large size of the acceleration region can easily satisfy the Hillas criterion and radiative losses in the tenuous medium of the galactic halo are not as significant as they are in the disk. Moreover, investigations on the composition of the cosmic rays at the end of the spectrum (e.g. \citealt{Auger2017a}) seem to show a trend towards heavy elements in accordance with the predictions of \cite{anchordoqui1999}. Recent results obtained by the Pierre Auger Observatory also suggest a marginally significant excess of events in the direction of some starburst galaxies \citep{AugerStarburst2018}.  However, \cite{romero2018} have investigated the case of NGC 253 using the new available information on the wind and its mass load obtained with the help of ALMA observatory \citep{bolatto2013} along with radio and X-ray observations and concluded that even under the most optimistic assumptions the superwind of this starburst \emph{cannot} accelerate cosmic rays beyond energies of $\sim 10^{16}$ eV for protons and  $\sim 4\times 10^{17}$ eV for iron nuclei. These results are in accordance with independent estimates by \cite{bustard2017}. Acceleration up to 100 EeV would require completely unrealistic magnetic fields outside this galaxy. In light of such a situation \cite{romero2018} suggested that ultra high-energy cosmic-ray acceleration in NGC 253 would still be possible if it is achieved by compact objects, either in a starved supermassive black hole at the galactic centre or in young pulsars in the disk. 

In this paper we shall explore the first of these possibilities, namely, we investigate the feasibility of cosmic-ray acceleration up to extreme energies in a putative hidden low-luminosity Active Galactic Nucleus (AGN) of NGC 253. With such an aim, we shall model the typical environment of such a nucleus. This involves a hot accretion flow which supplies matter to the vicinity of the central black hole and a magnetosphere where particles can be accelerated under certain conditions. As a by-product, we shall estimate the gamma-ray contribution of the AGN to the overall high-energy radiation emitted by the galaxy. 

The structure of the paper is as follows. In Section \ref{sec:NGC253} we summarise the main characteristics of the sources in the central region of NGC 253 that have been proposed to harbour a supermassive black hole. In Section \ref{sec:RIAF} we summarise the main characteristics of the accretion model we assume for these sources. In Section \ref{sec:gap} we describe the black hole dynamo mechanism and the conditions under which it is able to accelerate cosmic rays to very high energies. In Section \ref{sec:application} we apply the accretion flow model to the different nucleus candidates and probe the black hole dynamo mechanism. In Section \ref{sec:discussion} we discuss the implication of these results in the context of cosmic rays and gamma rays. We end the article presenting a summary in Section \ref{sec:summary}.

\section{The nearby starburst NGC 253 and its central region}
\label{sec:NGC253}

NGC 253 is a nearby edge-on starburst galaxy located in the Sculptor group ($\alpha \sim 00^{\rm h} 47^{\rm m},\delta \sim -25^{\rm o} ~17'$) at a distance of $3.5 \pm 0.2 ~ {\rm Mpc}$ \citep{rekola2005}. Along with M82, NGC 253 is the best studied starburst galaxy, and it has been detected at all wavelengths, from radio to high-energy gamma rays. The starburst of NGC 253 is believed to be fed by a $6$ kpc bar that funnels gas into the nucleus \citep{engelbracht1998}. A high number of neighboring galaxies contain both starbursts and active galactic nuclei, and it is not clear what is the physical connection between them \citep{levenson2001}. In the case of NGC 253, it is unknown whether an AGN coexists with the starburst or not. The physical properties and the nature of its nucleus are far from clear and have been discussed in the literature for many years. The central region of the galaxy is heavily obscured by gas and dust, and the effects of strong stellar winds affect kinematic studies. Historically, the most plausible nucleus candidate considered was the strongest compact radio source in the central region ($21~{\rm mJy}$ at $1.3$ cm), named TH2 after \cite{turner1985}. However, later studies did not find any counterpart in the infrared, optical or X-rays for this source. This lack of detection at other wavelengths led \cite{fernandez-ontiveros2009} to state that if TH2 holds a black hole, it has to be in a dormant state; they proposed that it could be a scaled-up version of Sgr A*.

Another galactic nucleus candidate proposed in the literature is the strong hard X-ray source X-1 \citep{weaver2002}. \textit{Chandra} observations show that this source is heavily absorbed ($N_H=7.5\times 10^{23}~{\rm cm}^{-2}$) and presents a very low $F_{0.5-2~{\rm keV}}/F_{2-10~{\rm keV}}$ ratio ($\sim 10^{-3}$). Based on this, \cite{mullersanchez2010} hypothesised that this source might be a hidden Low Luminosity Active Galactic Nuclei (LLAGN), similar to, but weaker than, the one found in NGC 4945 \citep{marconi2000}. They estimated an intrinsic $2-10$ keV luminosity of $\sim 10^{40}$ erg s$^{-1}$. 

Recently, based on near-IR observations, \cite{gunthardt2015} proposed a third nucleus candidate: the IR peak known as IRC. This source is coincident with the radio source TH7 \citep{ulvestad1997}, with the second strongest X-ray source, X-2, and with a massive star cluster of $1.4\times 10^7~{\rm M_\odot}$. They proposed that such a cluster can hide a $\sim 10^6~{\rm M_\odot}$ black hole. 

Whichever of these sources is the true galactic nucleus, the corresponding black hole must be in a low-activity state and hence it would be powered by a Radiatively Inefficient Accretion Flow (RIAF, see \citealt{yuan2014} for a recent review). In what follows  we shall assume this low-luminosity AGN scenario and we shall investigate the potential impact of the presence of a starved central black hole on cosmic ray acceleration and gamma-ray production in NGC 253.

\section{Accretion flow model}
\label{sec:RIAF}

We assume that the accretion flow around the supermassive black hole is in the form of a RIAF. We follow the standard semi-analytical modelling of these systems in the quiescent state. The standard model considers a steady axisymmetric hot, geometrically thick, optically thin, two-temperature flow with a low radiative efficiency \citep{narayan1997,yuan2000}. It is also well established that RIAFs present strong magnetocentrifugal-driven winds \citep{narayan1995,stone1999,yuan2012}.  To take into account the mass-loss of this wind we follow \cite{blandford1999} and make the accretion rate to depend on the radius via the parameter $s$ as:
\begin{equation}
    \dot{M}=\dot{M}_{\rm out} \Bigg ( \frac{R}{R_{\rm out}} \Bigg )^s,
\end{equation}
where $R_{\rm out}$ and $\dot{M}_{\rm out}$ are the outer radius of the flow and the outer accretion rate, respectively. Both numerical simulations \citep{stone1999,yuan2012} and observational studies \citep{quataert2000,wang2013} suggest that $0.1\lesssim s \lesssim 1$. The original works on radiatively inefficient ADAFs considered $\delta$, the energy released by turbulence that directly heats electrons, to be small ($\delta\sim 10^{-3})$ \citep{narayan1995}, but later works argued that this parameter is much higher, more likely lying in the range $0.1\lesssim \delta \lesssim 0.5$ (e.g. \citealt{quataert1999,sharma2007}). The other parameters that determine the hydrodynamic structure of the flow are the viscosity parameter $\alpha$, the gas pressure to magnetic pressure ratio $\beta$, and the black hole mass $M_{\rm BH}$. Given the poor observational constraints on the sources, we conservatively adopt standard values for most of the parameters, namely $\alpha=0.1$, $\beta=1$ (equipartition), $\delta =0.1$, and $s=0.3$. We allow the black hole mass to vary between $10^6$ and $10^7$ solar masses, and we adjust the accretion rate to fit the observational data. Given the parameters above, obtaining the global hydrodynamical solution is a two-point boundary value problem. We solve the equations numerically using the shooting method with appropriate boundary conditions, as outlined in \citet{yuan2000}. The flow extends from the Schwarzschild radius $R_{\rm S}$ up to $R_{\rm out}=10^4~R_{\rm schw}$, where the temperatures of ions and electrons are $T_{\rm out,i}=0.2 T_{\rm vir}$ and $T_{\rm out,e}=0.19 T_{\rm vir}$, respectively, and the Mach number is $\lambda \equiv v/c_{\rm s} = 0.2$; the virial temperature is given by $T_{\rm vir}=3.6 \times 10^{12}(R_{\rm schw}/R)~{\rm K}$. Once we obtain the density, pressure, magnetic field, and velocity of the flow, and the temperatures of both electrons and ions at each radius, we calculate the spectrum of the emission solving the radiative transfer equation via the ``plane-parallel rays'' method with boundary conditions for a non-illuminated atmosphere (\citealt{mihalas1978}, see \citealt{ozel2000,yuan2003} for more details). The emission processes considered for the thermal electrons are synchrotron emission, Bremsstrahlung (free-free) radiation, and inverse Compton up-scattering of the seed photons produced by those two processes.

\subsection{Jet component}
\label{subsec:jet}

We also allow the possibility that a small fraction of the matter that reaches the black hole is launched into a collimated outflow. We consider a simple one-zone leptonic model in which the jet starts magnetically dominated and at some distance from the hole develops internal shocks. A fraction $q_{\rm rel}<1$ of the jet power goes to non-thermal electrons, which emit synchrotron radiation (for more details see e.g. \citealt{boschramon2006,vila2010}). Following the ``disk-jet symbiosis'' hypothesis of \citet{falcke1995} (see also \citealt{kording2006}), the total power of the jet is a fraction of the accretion power,
\begin{equation}
    L_{\rm jet}=q_{\rm jet}\dot{M}(R_{\rm schw})c^2,
\end{equation}
where $q_{\rm jet}<1$. The other parameters of the model are the jet bulk Lorentz factor $\Gamma_{\rm jet}$, the viewing angle $i$, the distance from the black hole to the base of the acceleration region $z_{\rm acc}$, the opening angle of the jet $\theta_{\rm jet}$, the minimum Lorentz factor of the non-thermal leptons $\gamma_{\rm min}$, and the particle spectral index $\alpha$. We choose standard values used for AGNs \citep[see e.g. ][]{spada2001}: $\Gamma_{\rm jet}=2.3$, $\theta_{\rm jet}=0.1~{\rm rad}$, and $i=30\deg$, $\gamma_{\rm min}=2.0$, and we adjust the remaining parameters to fit the data when required.

\section{Acceleration by the black hole dynamo mechanism}
\label{sec:gap}

General relativity (GR) magnetohydrodynamic (MHD) simulations of accretion flows around rotating black holes show that a magnetically dominated funnel forms above the polar region of the black hole. Provided the force-free condition is fulfilled in this region, the rotational energy of the black hole can be extracted as an outwards Poynting flux via the Blandford-Znajek (BZ) mechanism \citep{blandford1977}. The initial poloidal field lines are twisted in the ergosphere of the black hole and develop a toroidal component.  Plasma with negative energy falls into the hole decreasing its energy and a magnetically dominated flux is evacuated along the rotation axis. The total power of this flux is $L_{\rm BZ}\sim c (B^2/8\pi) \pi r_{\rm g}^2 \propto B^2  r_{\rm g}^2 $, where $ r_{\rm g}=GM/c^2$ is the gravitational radius of the black hole.  In convenient units and introducing the black hole spin parameter $a_*$ this can be written as:
\begin{equation}
L_{\rm BZ} \approx 10^{46}\,\left(\frac{B}{10^4 {\rm G}}\right)^2 \left(\frac{M}{10^9 {\rm M_\odot}}\right)^2a_*^2\,\,\rm{erg\,s}^{-1}.
\label{eq:BZ_approx_power_2}
\end{equation}

\begin{figure}
    \centering
    \includegraphics[width=0.8\linewidth]{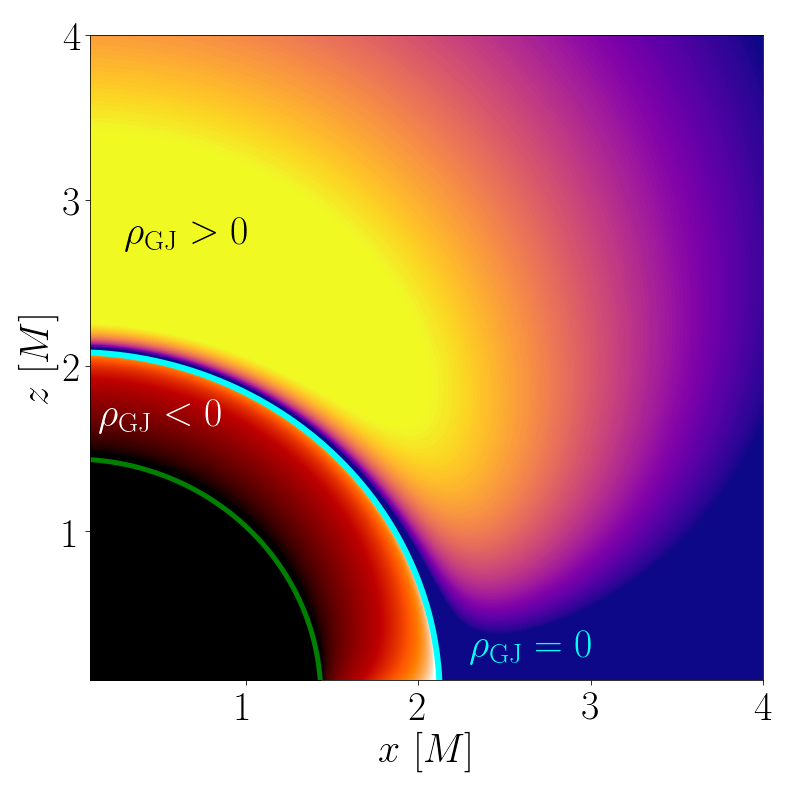}
    \caption{Goldreich-Julian charge density colour map for a black hole magnetosphere under the split-monopole magnetic field configuration. The green line shows the black hole event horizon and the cyan line shows the surface where $\rho_{\rm GJ}=0$. Adapted from \citet{ptitsyna2016}. The plot is only schematic and the colour map scale is arbitrary.}
    \label{fig:BHmagnetosphere}
\end{figure}

Since the plasma of the accretion flows cannot enter into the funnel because of the magneto-centrifugal barrier, another mechanism must be the responsible of providing the charge density required for a force-free plasma, namely the Goldreich-Julian charge density $\rho_{\rm GJ}$ \citep{goldreich1969}. One possibility to fill the magnetosphere is that MeV photons produced in the accretion flow collide between themselves and decay into electron-positron pairs in the funnel. However, if the luminosity of the disk is too low, as it is the case in RIAFs, the number of MeV photons could not be high enough to provide the magnetosphere with the Goldreich-Julian charge density. This lack of charges in some regions of the magnetosphere allows the formation of an unscreened electrostatic potential gap. The formation of this gap is located more likely near the region where $\rho_{\rm GJ}\sim 0$ (see Figure \ref{fig:BHmagnetosphere}).

For a nearly maximally rotating black hole, the potential drop in the gap can be estimated as:
\begin{equation}
    \Delta V \sim 4.5 \times 10^{17}~ M_6~B_4~\Big ( h/r_{\rm g} \Big )^2~{\rm V},
    \label{eq:deltaV}
\end{equation}
where $M_6 = M/10^6 {\rm M_\odot}$, $B_4=B/10^4 {\rm G}$, $h$ is the gap height, and $r_{\rm g}$ is the gravitational radius as before \citep{znajek1978, levinson2000}. The magnetic field is expected to be in equipartition with the accreting matter \citep{boldt1999}. This assumption leads to
\begin{equation}
    B_4 \approx 61 ~\Big ( \dot{m}/M_6 \Big )^{1/2},
\end{equation}
where $\dot{m}$ is the accretion rate at the horizon in Eddington units.

Thus, in principle, a charged nucleus of atomic number $Z$ entering into the gap can be accelerated up to very high energies,
\begin{equation}
    E_{\rm max} = Ze\Delta V \sim 3.0 \times 10^{19} ~ Z ~\dot{m}^{1/2}M_6^{1/2}\Big( h/r_{\rm g} \Big)^2~{\rm eV}.
\end{equation}
A strong limiting condition for this acceleration mechanism is that the gap height must be of the order of $r_{\rm g}$ for the process to be efficient. If electrons or positrons are also accelerated in the gap, they may close the gap or limit its extension via the emission of gamma rays through inverse Compton with a background photon field.
These gamma rays would collide with the same soft photon field giving rise to electron-positron pairs, which in turn emit more gamma rays. If this process is efficient enough, an electromagnetic cascade develops and the pairs created screen the electrostatic potential drop, thus inhibiting the acceleration. To account for the possibility of gap closure, we impose the condition that the multiplicity of pairs $\mathcal{M}$ (number of pairs created per lepton) should not be larger than 1 \citep{hirotani2016}.

The equation that governs the energy evolution of a lepton in the gap is
\begin{equation}
    \frac{dE}{dl} = \frac{e\Delta V}{h} - \frac{1}{c}\frac{dE}{dt} \Bigg |_{\rm IC} -
    \frac{1}{c}\frac{dE}{dt} \Bigg |_{\rm curv},
    \label{eq:energy}
\end{equation}
where the first term on the right hand side accounts for the acceleration, and the second and third terms are the energy losses by inverse Compton and curvature radiation, respectively. The curvature losses depend on the curvature radius, that we choose as $R_{\rm c} \sim r_{\rm g}$ \citep{ford2018}. For an electron of energy $E$, the number of pairs created via the inverse Compton process is
\begin{equation}
    \mathcal{M}_{\rm IC} = \int \frac{dN_{\rm IC}}{d\varepsilon_\gamma}(\varepsilon_\gamma) \Big [1-e^{-\tau_{\gamma \gamma}(\varepsilon_\gamma)} \Big ]d\varepsilon_\gamma,
    \label{eq:M_IC}
\end{equation}
where
\begin{equation}
    \frac{dN_{\rm IC}}{d\varepsilon_\gamma}(\varepsilon_\gamma) = \frac{h}{c}\frac{dN_{\rm IC}}{dtd\varepsilon_\gamma}(\varepsilon_\gamma),
    \label{eq:dNdE_IC}
\end{equation}
and
\begin{equation}
    \tau_{\gamma \gamma} (\varepsilon_\gamma) = h \int d\varepsilon_{\rm ph} n_{\rm ph}(\varepsilon_{\rm ph}) \sigma_{\gamma \gamma}(\varepsilon_\gamma,\varepsilon_{\rm ph}).
    \label{eq:tau_gg}
\end{equation}
In Eqs. \ref{eq:dNdE_IC} and \ref{eq:tau_gg}, $dN_{\rm IC}/dtd\varepsilon_\gamma$ is the inverse Compton emissivity (e.g. \citealt{blumenthal1970}), $\sigma_{\gamma \gamma} (\varepsilon_\gamma,\varepsilon_{\rm ph})$ is the photo-pair creation cross section \citep{aharonian1983}, and $n_{\rm ph}(\varepsilon_{\rm ph})$ is the spectral density of external soft photons in the gap. In our scenarios, this photon field is provided by the emission of the RIAF\footnote{The infrared emission of the starburst also provides seed photons to the inverse Compton process but in regions close to the black hole the energy density of these photon fields is approximately ten orders of magnitude lower than those produced by the RIAF.} (see Section \ref{sec:RIAF}).
We locate the inner gap boundary at $z_{\rm gap} = 2 r_{\rm g}$ \citep{ford2018} and assume homogeneity along its extension. Then, we obtain iteratively $h$ in each scenario by solving equations \ref{eq:energy} and \ref{eq:M_IC} imposing the gap closure condition $\mathcal{M}_{\rm IC}=1$ \footnote{Curvature radiation could also produce gamma rays and subsequent pair creation, but this process is subdominant \citep{hirotani2016}.}.

\section{Application to NGC 253}
\label{sec:application}

\subsection{Accretion flow}

We shall model now the physical scenario around the different black hole candidates in order to investigate the particle acceleration by the black hole dynamo mechanism. The uncertainties about the nature of the emission in these objects are important. We apply the hot accretion flow model assuming that the emission observed is powered by the black hole. This could not be the case, of course, and this does not rule out the possibility that one of these sources harbours a dormant black hole not responsible for the emission. The values adopted for the different parameters are shown in Table \ref{tab:model}.

\subsubsection{TH2}
\label{sec:TH2}
\begin{figure}
    \centering
    \includegraphics[width=1.0\linewidth]{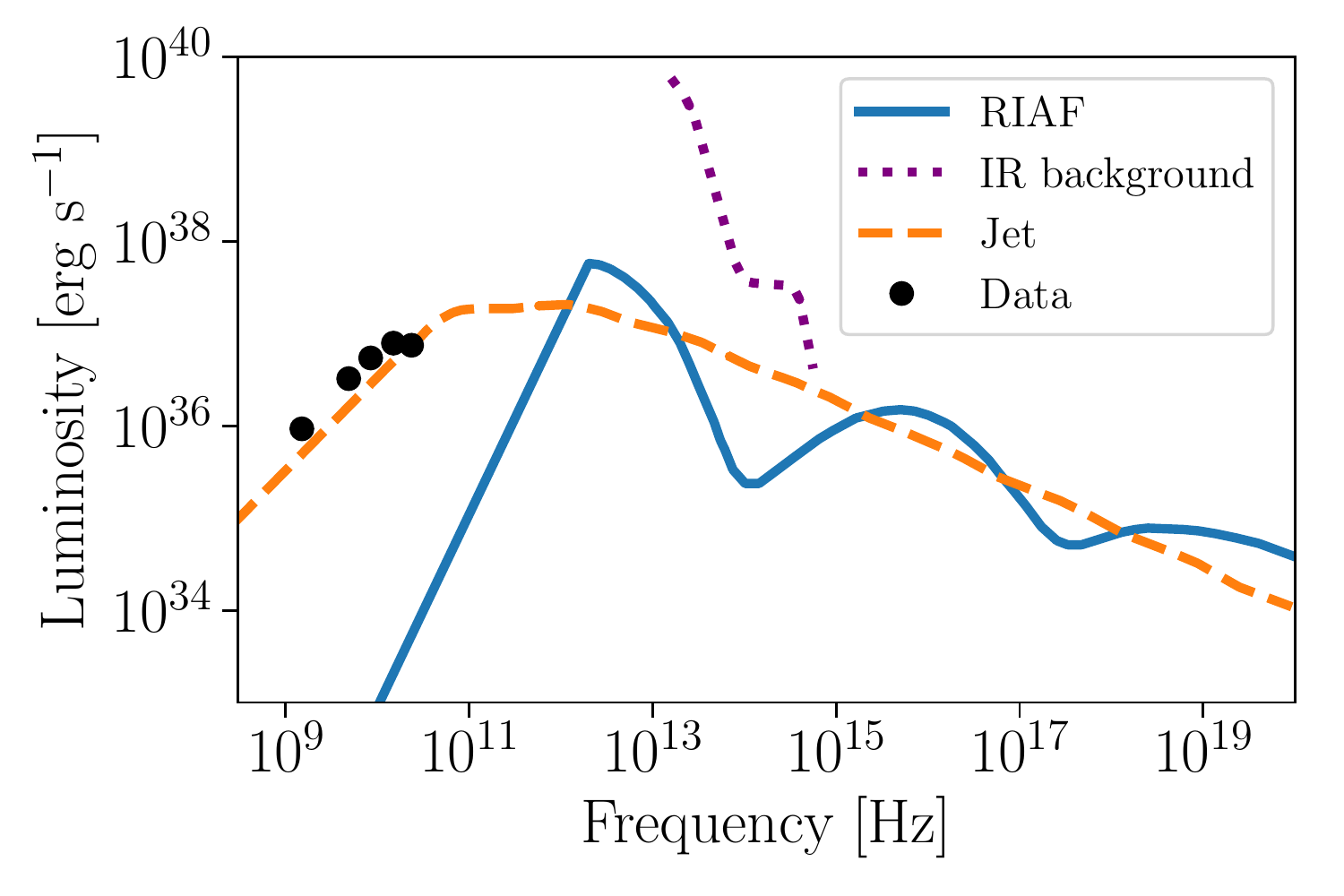}
    \caption{Spectral Energy Distribution (SED) of the hot accretion flow for the source TH2. The radio data are from \citet{ulvestad1997}, and the IR background limits are from \citet{fernandez-ontiveros2009}.}
    \label{fig:TH2}
\end{figure}

Despite TH2 being the strongest compact radio source in the nuclear region of NGC 253, it has no counterpart at other wavelengths. Hence, if this source corresponds to a supermassive black hole, this has to be accreting at very low rates. Given this starvation, we propose that the accretion flow onto the black hole is a RIAF similar to the one in Sgr A* \citep{narayan1998,yuan2003} but with a higher mass. \citet{fernandez-ontiveros2009} estimated a mass of $\approx 7 \times 10^6~{\rm M_\odot}$; given the uncertainties we adopt $M_{\rm BH}=10^7~ {\rm M_\odot}$. The thermal synchrotron emission of a RIAF around a supermassive black hole of this mass has its peak at submillimetre wavelengths and cannot be responsible for the detected centimetre flux. This emission, however, can be reproduced with the additional assumption that a weak jet is also present (see Section \ref{subsec:jet}). We take $q_{\rm jet}=10\%$, a spectral index for the non-thermal distribution in the jet of $\alpha=2.8$, and an outer accretion rate for the accretion flow of $\dot{M}_{\rm out}=1.1\times 10^{-4} \dot{M}_{\rm Edd}$. We assume that the acceleration of particles takes place at a distance of $z_{\rm acc} \approx 100r_{\rm g}$ from the black hole horizon. Figure \ref{fig:TH2} shows the SED of the RIAF+jet model that yields the best fit of the data. Though the RIAF emission is not seen, the jet power is linked to that of the accretion flow.

\subsubsection{X-1}
\label{sec:X1}

\begin{figure}
    \centering
    \includegraphics[width=1.0\linewidth]{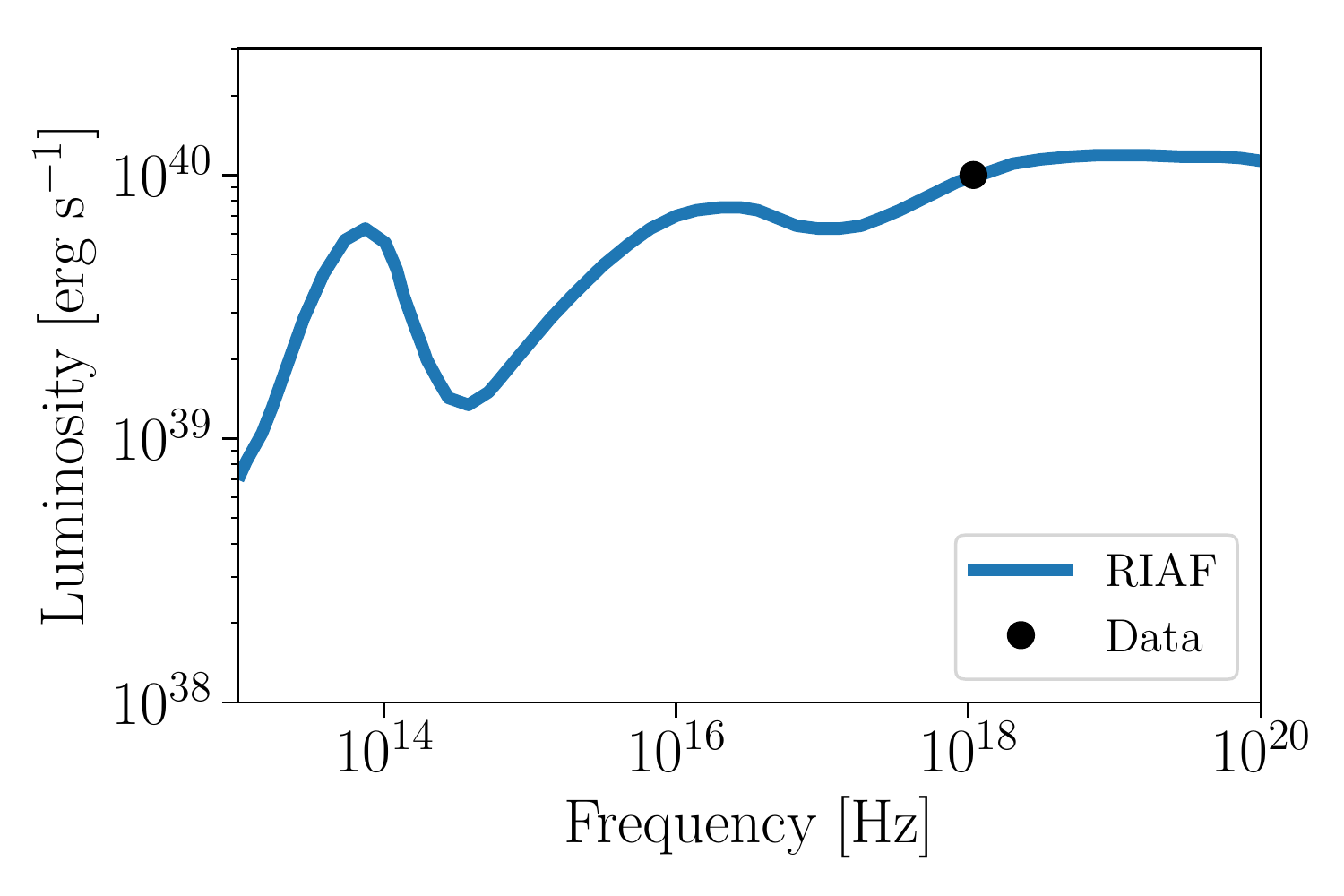}
    \caption{Spectral Energy Distribution (SED) of the hot accretion flow for the source X-1. The X-ray data point is from \citet{mullersanchez2010} .}
    \label{fig:X1}
\end{figure}

X-1 is a strong hard X-ray source in the nucleus of NGC 253. Early studies considered it to be the X-ray counterpart of TH2, thus reinforcing the AGN nature of the latter \citep{weaver2002}, but further reprocessing of the {\it Chandra} data by \citet{mullersanchez2010} demonstrated that both sources are not associated with each other, being separated by $\sim 1^{\prime \prime}$. Moreover, X-1 has no counterpart at other wavelengths, and it is only detected at energies $>2~{\rm keV}$. \cite{mullersanchez2010} stated that if X-1 is the true galactic nucleus, the simplest explanation is that it is a hidden LLAGN. Hidden AGNs are characterised by extremely high obscuration up to mid-infrared wavelengths, and they do not fit in the standard unified AGN model \citep{antonucci1993}. The prototype of this class of AGN is the Seyfert 2 galaxy NGC 4945 \citep{marconi2000}. \citet{mullersanchez2010} estimated the intrinsic luminosity of X-1 in the $2-10$~keV band to be $\sim10^{40}~{\rm erg ~s^{-1}}$. 
In 2013, simultaneous {\it Chandra} and {\it NuSTAR} observations of the central region of NGC 253 did not detect emission at the X-1 position \citep{lehmer2013}. This fact plus the upper limit imposed by {\it NuSTAR} ($L_{10-40~{\rm keV}} \lesssim 0.3 \times 10^{39}~{\rm erg~s^{-1}}$) seem to disfavour the AGN hypothesis. \citet{lehmer2013} suggested that if this source is a hidden AGN, it was in a low activity state during the 2013 observing campaign.  Adopting this assumption, we model the emission of X-1 as a RIAF with a mass of $M \approx 10^6~{\rm M_\odot}$ and an outer accretion rate of $\approx 1.5\times 10^{-2} \dot{M}_{\rm Edd}$. Figure \ref{fig:X1} shows the SED of the RIAF.

\subsubsection{IRC/TH7/X-2}

\begin{figure}
    \centering
    \includegraphics[width=1.0\linewidth]{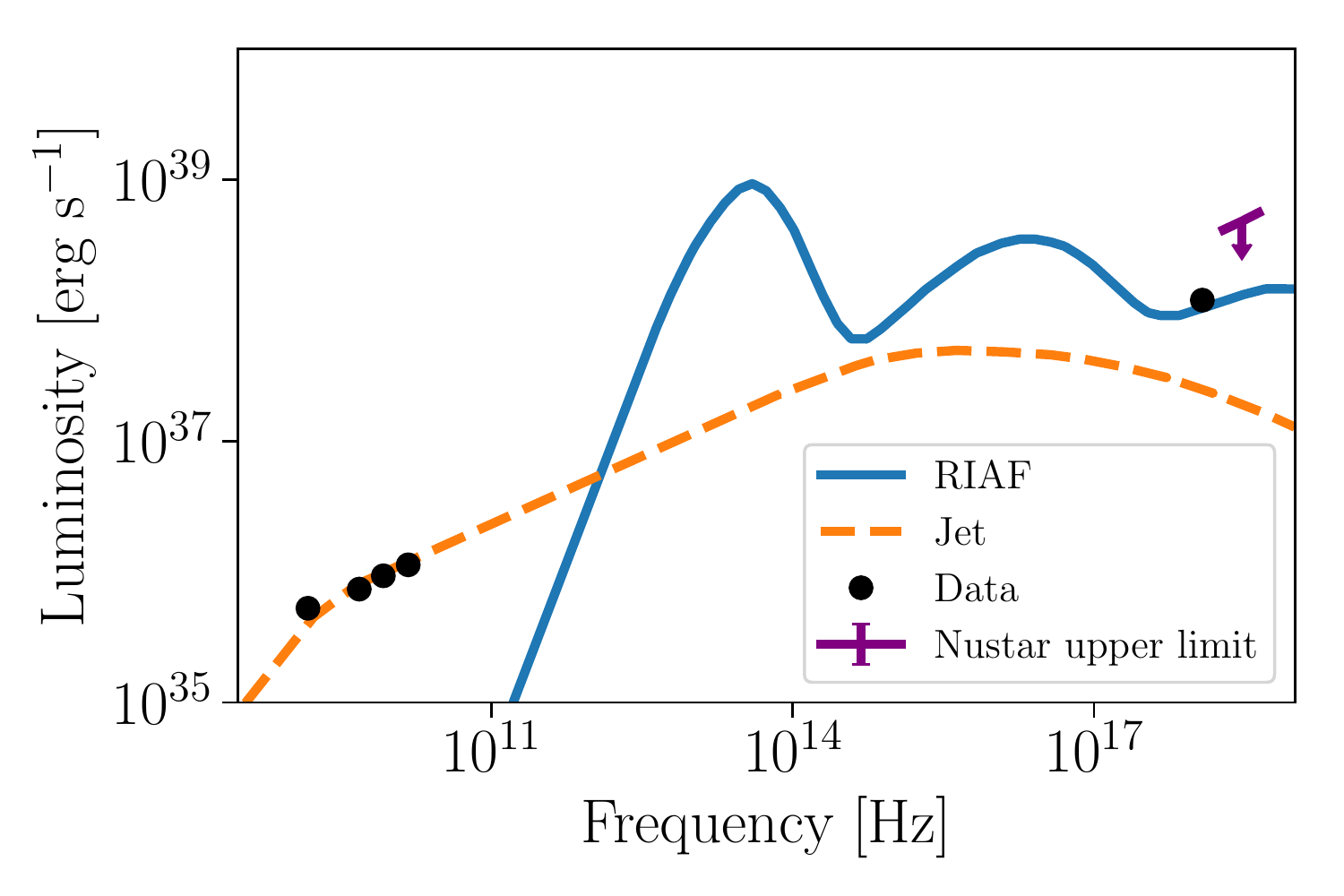}
    \caption{Spectral energy distribution for the accretion flow in IRC. The radio data are from \citet{ulvestad1997}, the $2-10~{\rm keV}$ data are from \citet{mullersanchez2010}, and the {\it NuSTAR} upper limits are from \citet{lehmer2013}.}
    \label{fig:IRC}
\end{figure}

IRC is the brightest near-infrared and mid-infrared source, as well as the most powerful soft X-ray source in the central region of NGC 253 (the source is dubbed X-2 in \citealt{mullersanchez2010}). It has a radio counterpart (TH7 in \citealt{ulvestad1997}), and it is coincident with a super stellar cluster (SSC) of $\sim 10^7~{\rm M_\odot}$. \citet{gunthardt2015} presented evidence suggesting that this source could be the true galactic nucleus and that in this case, the SSC might harbours a $\approx 10^6~{\rm M_\odot}$ low-luminosity accreting black hole. The luminosity in the {\it Chandra} band is $\sim 10^{38}~{\rm erg s^{-1}}$ \citep{mullersanchez2010} and can be reproduced by a RIAF with a black hole mass of $10^6~{\rm M_\odot}$ and an  outer accretion rate of $\dot{M}_{\rm out}\approx 2.5 \times 10^{-3}~ \dot{M}_{\rm Edd}$. Given the flatness of the radio emission, $S_\nu \sim \nu^{-0.8}$, the region in the jet that produces it should be located far from the black hole. We fit this data with a phenomenological jet model with $z_{\rm acc} \approx 2\times 10^4r_{\rm g}$, $\alpha=2.3$, and $q_{\rm jet}=8\%$. Figure \ref{fig:IRC} shows the SED for the RIAF+jet model. The IR luminosity of the source is very high ($\gtrsim 10^{42}$ erg s$^{-1}$ at the $K_s$ band) and it is very likely produced by the heated dust in the SSC and not related to the AGN; hence it is not shown in Figure \ref{fig:IRC}.

\begin{table*}
\centering
\caption{Parameters of the hot accretion flow and the jet (when necessary) for the galactic nucleus candidates.} 
\label{tab:model}
\setlength{\tabcolsep}{4.5pt}
\begin{tabular}{l|cc|cccc}
\hline 
~~~~~~~~~~~~~~~~~~~~~~~~~~~~~~& Accretion flow & & & Jet \\ [0.5ex]
\hline
\hline
Source      ~~~~~~~~~~~~~~     &  $M_{\rm BH}~[10^6~{\rm M_\odot}]$     &  $\dot{M}_{\rm acc}~[\dot{M}_{\rm Edd}]$ & $q_{\rm jet}$ & $\alpha$ & $z_{\rm acc}~[R_{\rm S}]$ \\ [0.5ex]
\hline

    TH2  & $10$   &  $1.1 \times 10^{-4}$   & $10 \%$     & $2.8$ &   $100$   \\ [0.5ex]
    X-1  & $1$    &  $1.5 \times 10^{-2}$   & $-$  & $-$ & $-$  \\ [0.5ex]
    IRC/TH7/X-2  & $1$    &  $2.5 \times 10^{-3}$  & $8 \%$      & $2.3$      &   $10^4$   \\ [0.5ex]
    \hline
\end{tabular}
 
\end{table*}

\subsection{Electrostatic gaps}

To probe the potential of the black hole dynamo mechanism to accelerate cosmic rays in this scenario, we must take into account the influence of the RIAF photon field on the gap extension. We impose the condition that an electron injected in the gap does not trigger an exponential growth of pairs via the emission of gamma rays (see Sec. \ref{sec:gap}). We take into account inverse Compton and curvature losses that limit the maximum energy of the leptons. From the sources considered, we find that even for the low-luminosity RIAF considered for TH2 the gap height is heavily reduced: $h \approx 0.01r_{\rm g}$. For a magnetic field $B \approx 600~{\rm G}$ as there is in its innermost region, the maximum energy that a charged nucleus can achieve is
\begin{equation}
    E_{\rm max} \lesssim 3 \times 10^{13}~ Z~{\rm eV}. \label{eq:E_max}
\end{equation}

\begin{figure}
    \centering
    \includegraphics[width=0.8\linewidth]{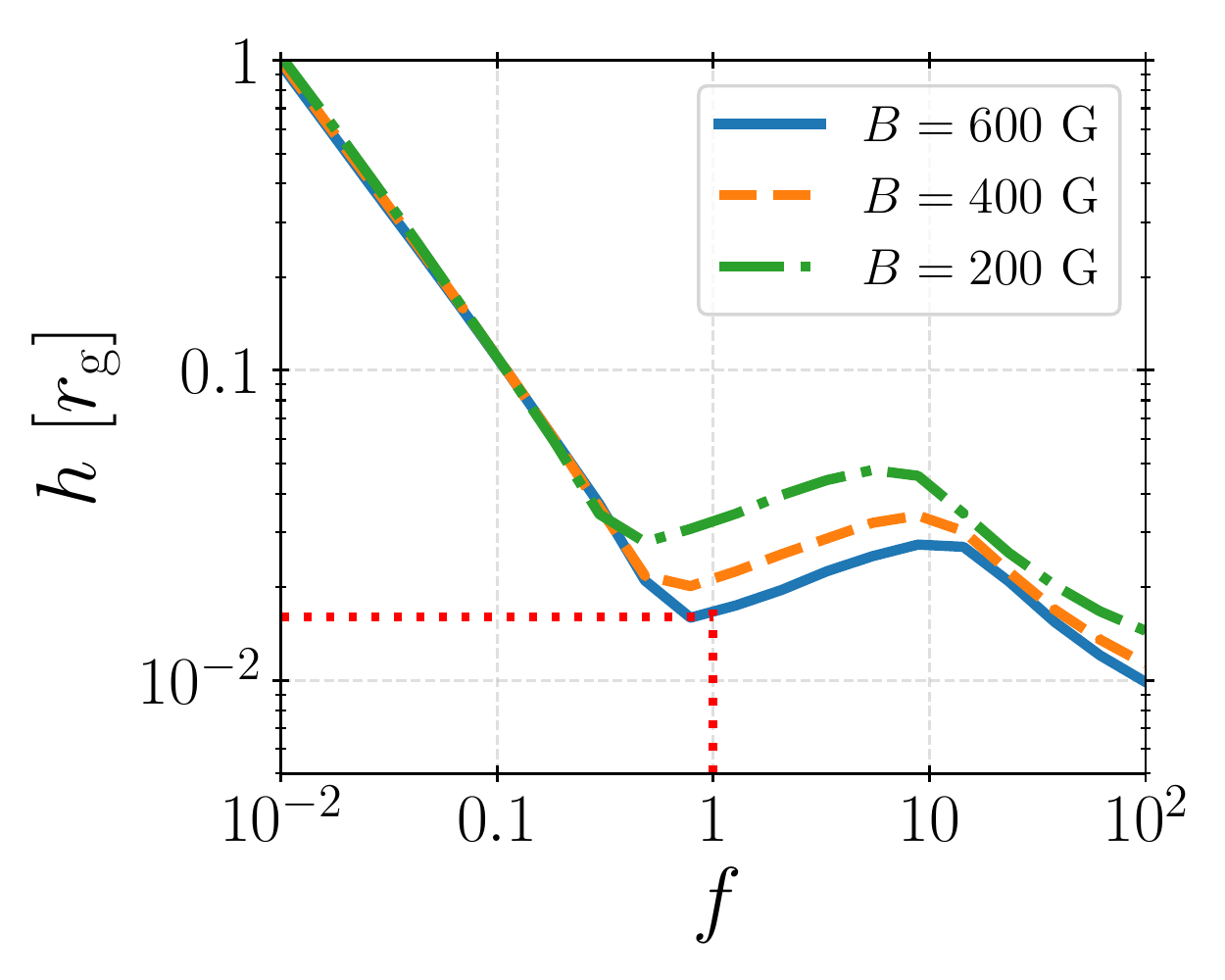}
    \caption{Gap height as a function of the parameter $f$ that parametrises the photon density, $n_{\rm ph} = f n_{\rm ph}^{\rm (TH2)}$, for different values of the magnetic field intensity. The dotted-red line shows the location in the plot of the parameters for the source TH2 (see Table \ref{tab:model}).}
    \label{fig:gap}
\end{figure}

For cosmic rays to be accelerated to higher energies in the electrostatic gap a lower RIAF luminosity is required. Despite the shape of the spectrum changes as the luminosity of the RIAF increases or decreases, at low accretion rates the most relevant seed photons are those of the sub-millimeter peak. Hence, to study how the gap height is affected by the background radiation it is a good approximation to take the photon spectrum from the RIAF in TH2 and scale it to different situations. In figure \ref{fig:gap} we show the dependence of the gap height with the intensity of the background photon field parametrized by $f$ such that the photon density is $n_{\rm ph} = f n_{\rm ph}^{\rm (TH2)}$. The little bump in the plot is related to the change from the Thomson regime to the Klein-Nishina regime in the inverse Compton emission of the electron. We see that for the gap height to be $\sim 1$ a photon density approximately two orders of magnitude lower would be required. In the case of TH2, it could be in principle that such a fainter accretion flow is present but, since the emission of the jet is linked to that of the RIAF by the jet-disk symbiosis, the jet power should be much higher than $> 10\%$ of the accretion power or its emission should be strongly beamed, which is not expected in this scenario.
We conclude that if TH2 harbours a starved black hole that is powering the radio emission, the potential of the black hole dynamo mechanism to accelerate cosmic rays in this source is heavily limited because of the emission of the accretion flow.
The RIAFs we considered for X-1 and IRC are both quite brighter than the one in TH2. So, if any of these sources is related to the true active nucleus, the gap would be even shorter and the acceleration less efficient. This rules out the putative LLAGN as a viable alternative for ultra high-energy cosmic-ray acceleration in NGC 253. 

There is still an open window for an extremely dormant black hole, either coincident with these sources or not, which cannot be completely ruled out until the true nature of the galactic nucleus in NGC 253 is clarified.

\section{Discussion}
\label{sec:discussion}

\subsection{Cosmic rays}

The presence of a hidden AGN in NGC 253 cannot explain the origin of ultra high-energy cosmic rays (UHECRs) in this source. According to Eq. \ref{eq:E_max}, the maximum energy attainable by iron nuclei entering into the electrostatic gap of the black hole magnetosphere would be $\sim 2\times 10^{15}$ eV. This is even less than what \cite{romero2018} estimated for particle acceleration in the superwind. If NGC 253 and other starbursts are confirmed as sources of UHECRs of high metalicity, then the cosmic rays should probably be originated in compact objects located in the star-forming region. Amongst the candidates we can mention engine-driven supernovae \citep{Zhang2019}, a recent tidal disruption event \citep{Guepin2018}, magnetars \citep{Singh2004}, and gamma-ray bursts \citep{Waxman2006}.

\subsection{Different contributions to gamma rays}

NGC 253 has been detected at high-energy (HE) gamma rays by the LAT instrument onboard of the \emph{Fermi} satellite \citep{abdo2010b} and at very high energies (VHE) by HESS \citep{acero2009}. The overall gamma-ray emission is analysed by \cite{Abramowski2012}. At a distance of 3.5 Mpc the integrated flux above 200 MeV yields a luminosity of $L(E>200 \; {\rm MeV})\sim 7.8 \times 10^{39}$ erg s$^{-1}$. The HE spectrum is well-fitted as a power-law of index $\Gamma=2.24 \pm 0.14_{\rm stat} \pm 0.03_{\rm sys}$. The VHE spectrum observed by HESS is also a power law with a similar index of $\Gamma=2.14 \pm 0.18_{\rm stat} \pm 0.30_{\rm sys}$. Both spectra are compatible with an overall spectrum fitted by a power law of index $\Gamma=2.34\pm 0.03$. However, a slight break cannot be ruled out with a lightly harder spectrum at higher energies. 

The observations with the best resolution in gamma rays are those of HESS. The detection is consistent with the galactic centre and with a source whose extension is of less than $2.'4$ at $3\sigma$ confidence level \citep{Abramowski2012}.  Since the extension of the starburst region is $\sim0.'4 \times 1.'0$, there is some room for contributions from the base of the superwind as suggested by \cite{romero2018}. GeV gamma rays might also be produced in the superwind, but \textsl{Fermi} resolution does not allow to disentangle this radiation if present. On the other hand, the observed IR-radio correlation of star-forming galaxies has led to the idea that cosmic-ray electrons accelerated by supernovae are responsible for the radio emission of these sources. In addition, cosmic-ray  protons produced also by supernovae might be responsible for the gamma-ray emission through inelastic collisions with ambient gas \citep{anchordoqui1999,domingo2005,rephaeli2010,paglione2012,yoasthull2013}. Other sources of locally accelerated protons, such as the stellar winds of massive stars, might also contribute to the overall gamma-ray luminosity (e.g., \citealt{romero2003b,yoasthull2014b}).

\subsection{Contribution from a hidden LLAGN to the gamma-ray emission}
\label{subsec:gamma}

NGC 4945, the prototype of hidden AGN, is one of the few radio-quiet AGNs detected by the Fermi-LAT telescope \citep{abdo2010}. Interestingly, despite this galaxy has also a starburst in its central region, recent evidence suggests that the high-energy emission could be powered by the accretion flow around the central supermassive black hole \citep{wojaczynski2017}. Motivated by this, we explore here what would be the maximum contribution to the gamma-ray emission expectable from a similar but weaker hidden AGN in NGC 253, as the one that might be responsible for the emission in X-1.

\begin{figure}
    \centering
    \includegraphics[width=1.0\linewidth]{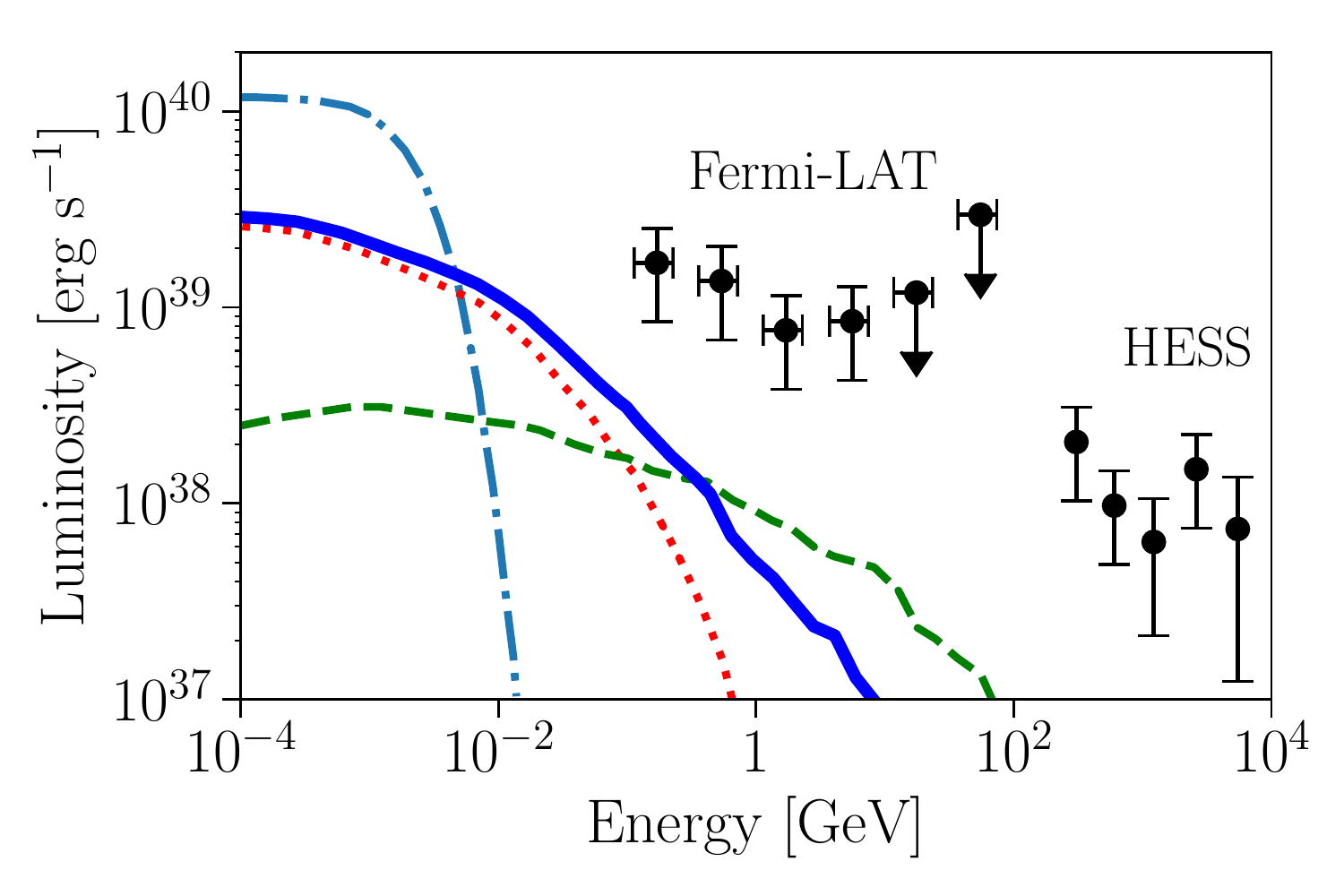}
    \caption{Leptonic scenario for the gamma-ray emission associated with a hidden-LLAGN in X-1 during a high-activity state (see Table \ref{tab:model}). The dotted line is synchrotron emission, the dashed line is inverse Compton emission, and the dot-dashed line is the thermal RIAF emission. The solid line is the total nonthermal emission including photo-pair absorption. The data points are from the Fermi-LAT \citep{abdo2010b} and HESS \citep{acero2009}}
    \label{fig:X1_g}
\end{figure}

Since RIAFs are plasmas in the collisionless regime, it is not clear whether particles in them are thermal or not \citep{mahadevan1997}. Moreover, at least in some LLAGNs, there is evidence that points to the presence of a non-thermal component \citep{inoue2018, yuan2003}. We explore the most favorable scenario for gamma-ray production in a hot accretion flow around a $\sim 10^6~{\rm M_\odot}$ black hole with a luminosity $\sim 10^{40}~{\rm erg~s^{-1}}$ in the {\it Chandra} band (see Sec. \ref{sec:X1}). We assume that a fraction of the energy density of both electrons and ions follows a non-thermal distribution of spectral index $p$. Some plausible acceleration mechanisms in these plasmas are turbulent magnetic reconnection (see \citealt{hoshino2012} for a review), stochastic acceleration \citep{dermer1996}, or diffusive shock acceleration \citep[e.g, ][]{drury1983,blandford1987}. Electrons cool locally by synchrotron radiation, so their cooled spectrum is steeper with an increased index of $p+1$ ($N(E)\propto E^{-(p+1)}$). On the other hand, ions do not cool efficiently and are advected toward the hole; as a consequence, they preserve the original spectral index. We explore a leptonic scenario where $10\%$ of the electrons (this fraction cannot be much higher or the low energy emission would be overestimated) follows a non-thermal distribution with $p=2$, and a hadronic scenario where all ions are non-thermal, following a harder power-law distribution of index $p=1.5$, and we estimate the associated gamma-ray emission in both cases. Electrons emit by synchrotron radiation and inverse Compton radiation, while ions emit synchrotron and also produce pion-decay gamma rays via $pp$ interactions; however, the latter process produces photons at energies where they are completely self-absorbed. The detailed description of the model is not the aim of this work and it will be presented in a forthcoming paper (Guti\'errez et al. 2020, in preparation). Figure \ref{fig:X1_g} shows the high-energy SED for the parameters above in the leptonic scenario. The electrons are able to produce $\approx 10 \%$ of the GeV emission detected from NGC 253. Photons with higher energies will be internally absorbed.
\begin{figure}
    \centering
    \includegraphics[width=1.0\linewidth]{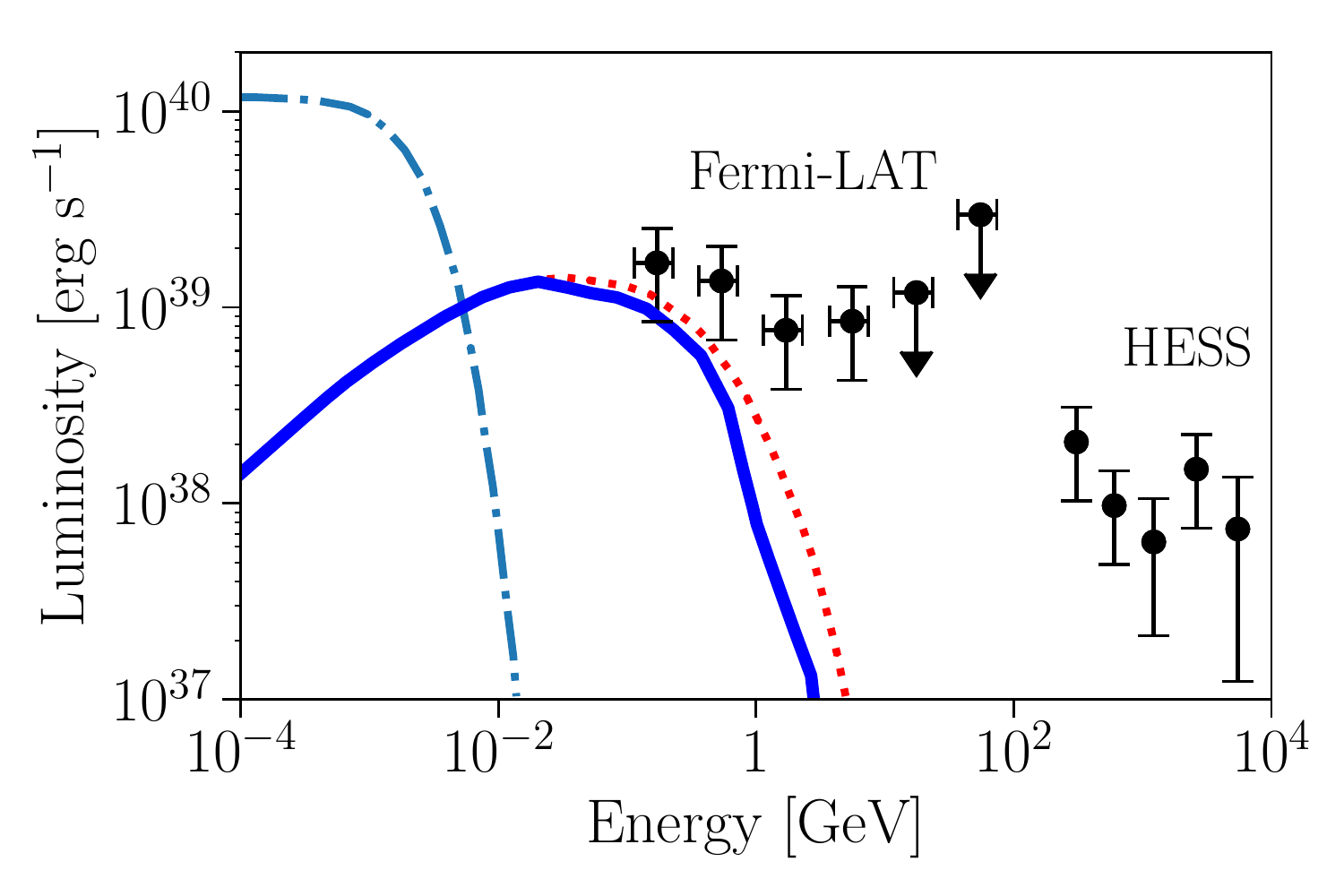}
    \caption{Hadronic scenario for the gamma-ray emission associated with a hidden-LLAGN in X-1 during a high-activity state (see Table \ref{tab:model}). The dotted line is synchrotron emission internally produced and the solid line is the final outgoing nonthermal emission including photo-pair absorption. The data points are from the Fermi-LAT \citep{abdo2010b} and HESS \citep{acero2009}}
    \label{fig:X1_g2}
\end{figure}
Figure \ref{fig:X1_g2} shows the hadronic case; given the high magnetic fields, if an efficient acceleration mechanism occurs protons are able to emit GeV synchrotron photons with a luminosity comparable to the total emission detected from NGC 253 at this band.

The above predictions should be taken as upper limits for the contribution of a central active galactic nucleus to the high-energy emission of NGC 253, since we are adopting the most favorable sets of values for the parameters. Nevertheless, given the low contribution of the RIAF in the GeV band and the lack of emission in the HESS energy range, the non-thermal processes in the RIAF are unlikely to ease the difficulties found in the joint fit of radio and gamma rays in the context of the calorimeter model for NGC 253 \citep{yoasthull2014a}.

To constrain the content of high-energy hadrons a self-consistent study of the neutrino production should be performed. We will calculate the neutrino emission self-consistently with the gamma radiation in a subsequent work.
Neutrinos might provide also a powerful test to investigate the relative contribution of starburst and AGNs to the high-energy regime in galaxies where the two phenomena coexist; both starburst galaxies and LLAGN have been considered as possible sources of very high-energy neutrinos. Moreover, recently some authors have proposed that RIAFs might play a role in the neutrino emission of LLAGNs (e.g. \citealt{kimura2015,kimura2019}).

\section{Summary and conclusions}
\label{sec:summary}

We have investigated the feasibility of cosmic-ray acceleration via the black hole dynamo mechanism in a putative active nucleus of NGC 253. We have considered the three most plausible candidates in the galactic centre region for a $10^{6-7}~{\rm M_\odot}$ black hole. Whichever the black hole is, it should be accreting at low rates and hence it would be powered by a RIAF. We modelled the electromagnetic emission of these flows to reproduce the observational data; for two sources, TH2 and IRC, we also assumed the presence of a small jet component to account for the radio emission. We have studied the viability of cosmic-ray acceleration to very high energies by an electrostatic potential gap in the polar region of the black hole magnetosphere. Charges accelerated in the gap emit gamma rays that collide with the soft photons from the RIAF and pair-create, thus closing the gap. Taking into account this effect, we have found that even in the least luminous of the three flows studied, that of TH2, the RIAF luminosity is high enough to limit heavily the extension of the gap, avoiding cosmic rays to be accelerated to energies higher than $10^{15}~{\rm eV}$. This effect rules out the AGN scenario as a candidate for UHECR acceleration in the case that the nucleus is powering the emission of one of the source candidates considered in the literature. There is still the possibility of an extremely weakly accreting nucleus, that is not seen in any way, to play some role in cosmic ray acceleration. To rule out this latter possibility it would be necessary either a clarification concerning the true nature of the NGC 253 nucleus, or a robust alternative explanation for the UHECR origin.

Finally, as a by-product of the modelling of the source X-1, we have estimated the contribution from non-thermal particles in the accretion flow to the overall gamma-ray emission of NGC 253. In the most favourable scenario, protons might produce a luminosity comparable to the one detected by the Fermi-LAT telescope at $\approx 1$ GeV.

\section*{Acknowledgements}

This work was supported by the Argentine agency CONICET (PIP 2014-00338), the National Agency for Scientific and Technological Promotion (PICT 2017-0898), and the Spanish Ministerio de Econom\'ia y Competitividad (MINECO/FEDER, UE) under grant AYA2016-76012-C3-1-P.




\bibliographystyle{mnras}
\bibliography{biblio} 







\bsp	
\label{lastpage}
\end{document}